\documentclass[twocolumn,showpacs,preprintnumbers,amsmath,amssymb]{revtex4}

\usepackage{amsbsy}
\usepackage{graphicx}

\begin{document}

\title{Cryptoferromagnetism in Superconductors with Broken Time-Reversal-Symmetry}

\author{N.~A.~Logoboy and E.~B.~Sonin}

\affiliation{Racah Institute of Physics, The Hebrew University of
Jerusalem, Jerusalem 91904, Israel}

\date{\today}

\begin{abstract}

The cryptoferromagnetic state (the state with intrinsic domain
structure) in superconducting ferromagnets subjected to external
magnetic field is studied theoretically. Ferromagnetism originates
either from electron spin or the intrinsic angular momentum of
Cooper pairs (chiral $p$-wave superconductors like $\mathrm{Sr_2 Ru
O_4}$).  The phase transitions towards the Meissner and the mixed
states are investigated, and the magnetic phase diagrams are
obtained. Cryptoferromagnetism, as a form coexistence of superconductivity and ferromagnetism, can be detected by observation of magnetization curves predicted in the present analysis.

\end{abstract}


\maketitle

In recent years numerous experimental evidences of
superconductivity-ferromagnetism coexistence  in various materials
were reported
\cite{Felner,Eskildsen,Luke,ishida,Saxena,Pfleiderer}. Two
types of such coexistence are possible: (i) The phase transitions to
the ferromagnetic and the superconducting (SC) states occurs at
different temperatures, so the coexistence starts below the lower
from the two transitions. Rutheno-cuprates \cite{Felner} belong to
this type: the superconductivity onset occurs at the temperature
much lower than the temperature of the magnetic transition. Normally
different elements of the crystal structure are responsible for
ferromagnetism and superconductivity, and spontaneous magnetization
(ferromagnetic order parameter) is related to spin. Later we call
them spin superconducting ferromagnets (spin SFs). (ii) The magnetic
and the SC transitions occur simultaneously. This can take place in
unconventional superconductors with triplet Cooper pairing. An
example of them is  strontium ruthenate $\mathrm{Sr_2 Ru O_4}$
\cite{Luke,ishida,mack}. The theory connects spontaneous
magnetization in this material not with spin but  with the orbital intrinsic
angular moment of the $p$-wave  Cooper  pair with the wave function
in the momentum space proportional to $p_x + ip_y$ (chiral p-wave
superconductivity). We shall call them orbital superconducting
ferromagnets (orbital SFs).

Whereas proof of superconductivity in SF materials is quite
straightforward, a clear-cut detection of the ferromagnetic order
parameter is much more problematic. The internal magnetic field is
screened out by the SC Meissner currents and can be present only
near sample borders and defects, in particular, domain walls (DWs).
This strongly suppresses the stray magnetic  fields around the
sample, which are most convincing evidence of ferromagnetism.
Especially worrying is situation with strontium ruthenate
$\mathrm{Sr_2 Ru O_4}$, where Kirtley {\em et al.} \cite{Kirtley}
could not detect any stray field from DWs or sample edges at all.
This is a challenge for the theory and for the very scenario of
chiral p-wave pairing. Difficulties with direct detection of
ferromagnetism coexisting with superconductivity lead to the
question  whether one may use the term {\em ferromagnetism} at all.
Indeed in the literature on unconventional superconductors
sometimes they prefer to tell about {\em superconductivity
with broken Time-Reversal Symmetry} (TRS). We used this name in the
title of the paper following this more cautious semantics though one
cannot imagine broken TRS without at least some features of
ferromagnetism (see further discussion).

Among possible explanations why  experimentalists cannot see stray
fields from DWs is  the presence of domain structure with a period
essentially smaller than a distance between a sample surface and a
probe used by experimentalists. There were some experimental
evidences of domains in SFs both in the spin \cite{LTSF} and the
orbital SF \cite{dom}. The theoretical investigations of the domain
structure in SFs were restricted with the case of zero external
magnetic field \cite{Krey,son02,Faure,Sonin2}. One must discern two
possible types of equilibrium domain structure. The first one is
well known for normal ferromagnets \cite{LL}. The domain structure
results from competition between the energy of DWs and the
magnetostatic energy of stray fields generated by the magnetic flux
exiting from the sample surface. The period of the structure depends
on the shape and the size of the sample going to infinity when the
sample size grows. One can call these domains {\em extrinsic
ferromagnetic domains}. Since in SFs  the Meissner effect expels the
magnetic field, it is impossible to benefit from decreasing the bulk
magnetostatic energy in comparison with the DW energy, and extrinsic
domains cannot appear at equilibrium \cite{son02}. But also long ago
there was known another type of domains, which decrease the bulk
magnetostatic energy at the expense of destroying the Meissner state
\cite{Krey,Faure,Sonin2}. The size of these domains is roughly of
the order of the London penetration depth $\lambda$ and does not
depend on either shape or size of the sample. Strictly speaking the
state with this domain structure at the macroscopic scales is not
ferromagnetic but antiferromagnetic  though with a rather large
period. We shall call such a state {\em cryptoferromagnetic}, the
term introduced by Anderson and Suhl \cite{Anderson} for another
model of ferromagnetism and superconductivity, in which crystal
anisotropy was neglected and spiral structure appeared instead of
domains.

In present publication we extend the theory of intrinsic domain structure (cryptoferromagnetic state) on nonzero external magnetic field and analyze competition of the cryptoferromagnetic state with the pure Meissner state and  the mixed state with vortices. This allows to obtain the full phase diagram of both spin and orbital SFs.
We demonstrate that the measurement of magnetization curves in various areas of the phase diagram can provide evidence of ferromagnetic or cryptoferromagnetic order in superconductors with broken TRS.

Let us consider a stripe domain structure with $180^{0}$ DWs  in a sample subjected to external magnetic field $\mathbf
{H}_{0}=(0,H_{0},0)$. The DWs are parallel to the $yz$-plane
separating domains with alternating magnetization $\mathbf{M}=(0,\pm
M_0,0)$ along the $+y$ or $-y$ direction. Since the $\mathbf
{H}_{0}$ orientation is preferable the width $d_\uparrow$ of domains
with the magnetization $\mathbf {M}$ parallel to $\mathbf {H}_{0}$
($\uparrow$-domains) exceeds the width $d_\downarrow$ of the domains
with $\mathbf {M}$ antiparallel to $\mathbf
{H}_{0}$ ($\downarrow$-domains). We restrict ourselves to the
simplest case when the London penetration length $\lambda$ exceeds
the DW thickness. Then the surface energy $\sigma$ and the internal
structure of DW is not affected by fields and currents at scales of
$\lambda$.

The Gibbs potential inside domains is
\begin{equation} \label{eq:free energy}
\mathcal {G}=\int d^3x \left(\frac{\mathbf {h}^{2}}{ 8\pi}+\frac{2\pi
\lambda^2}{c^2}\mathbf {j}^{2}-\mathbf {h}\cdot \mathbf
M-\frac{\mathbf {h}\cdot \mathbf
{H}_{0}}{ 4\pi}\right),
\end{equation}
where $\mathbf {h}$ is the magnetic field, and the electric current
$\mathbf{j}$ is connected with the magnetic field $\mathbf {h}$ via
the Maxwell equation $\mathbf{\nabla}\times \mathbf{h}=(4\pi/c)
\mathbf{j}$. Variation of the Gibbs potential yields the magnetic
field
$\mathbf{h}_{\uparrow,\downarrow}=(0,h_{\uparrow,\downarrow},0)$ in
the $\uparrow$-domains and $\downarrow$-domains:
\begin{equation} \label{eq:bh}
 h_{\uparrow,\downarrow}= (H_0\pm 4\pi M_0) \frac{\cosh {(x/\lambda-\xi_{\uparrow,\downarrow}})}{\cosh{\xi_{\uparrow,\downarrow}}},
\end{equation}
where  $x$ is the distance from the DW and
$\xi_{\uparrow,\downarrow}=d_{\uparrow,\downarrow}/2\lambda$ are
reduced domain widths.

Application of the Gibbs potential Eq.~(\ref{eq:free energy}) to
orbital SFs requires some comments. As shown in Ref. \cite{Braude},
for orbital ferromagnetism related to the intrinsic angular momentum
of Cooper pairs the spontaneous magnetization cannot be defined
unambiguously and therefore the Landau-Lifshitz theory of
ferromagnetism \cite{LL} based on this definition is not valid.
Nevertheless, interaction of magnetization currents inside narrow DW
with the magnetic field can be reduced to the expression looking
like the standard Zeeman energy $- \mathbf {h}\cdot \mathbf{M_0}$.
However, here $\mathbf{M_0}$ is not a magnetic moment inside the
domain but is defined so that $8\pi M_0$ would be the jump $8\pi
M_0$ of the magnetic field on the DW [see Eq.~(\ref{eq:bh})]. So
``magnetization'' $M_0$ is determined by the DW structure and cannot
be used for other phenomena connected with ferromagnetic ordering,
e.g., analyzing the magnon spectrum \cite{Braude}.

 Substituting Eq.~(\ref{eq:bh}) into
Eq.~(\ref{eq:free energy}), adding the surface energy $\sigma$ of
DWs, and averaging over the domain-structure period
$d=d_\uparrow+d_\downarrow$ we arrive to the following expression
for reduced energy density $\mathcal {E}=\mathcal {G}/2\pi
M^{2}_{0}V$ ($V$ is the sample volume):
\begin{eqnarray} \label{eq:Reduced Energy Density}
   \mathcal {E}
  =\frac{2w-(1+h_0)^2 \tanh \xi_\uparrow-(1-h_0)^2 \tanh \xi_\downarrow}{\xi_\uparrow+\xi_\downarrow}.
          \end{eqnarray}
Here $h_0=H_0/4\pi M_0$ and $w=\sigma/4\pi M^{2}_{0}\lambda$ are
dimensionless parameters. If $h_0=0$ and
$\xi_\uparrow=\xi_\downarrow$ Eq.~(\ref{eq:Reduced Energy Density})
coincides with the free energy density of Krey \cite{Krey}.

\begin{figure}
  \includegraphics[width=0.4\textwidth]{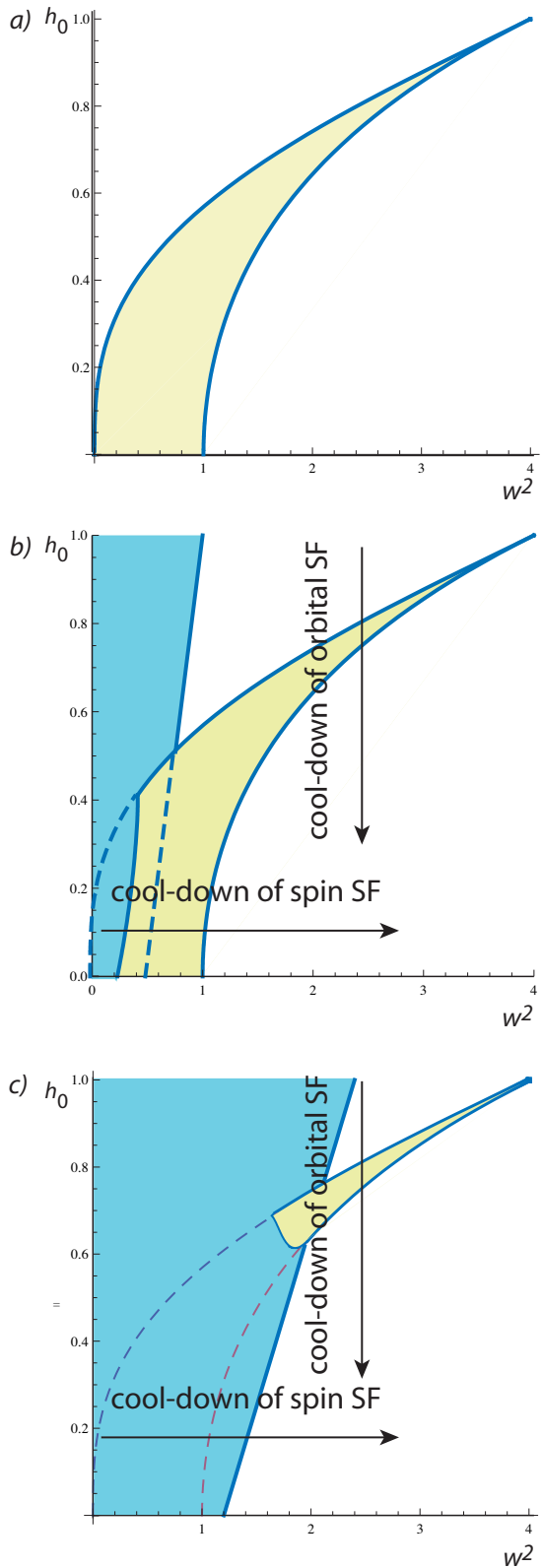}
\caption{ (color online) Phase diagram for various values of the
reduced lower critical field $h_{c1}$:  (a) $h_{c1}\to \infty $; (b)
$h_{c1}=2$; (c) $h_{c1}= 0.83$. The lighter (yellow) shaded area is
the cryptoferromagnetic state. The darker (blue) shaded area is the
mixed state. The rest is the Meissner state. The horizontal and the
vertical arrows show the processes of cool-down across the SC
critical temperature of spin and orbital SFs respectively.}
 \label{PhD}
\end{figure}

Minimization of energy density Eq.~(\ref{eq:Reduced Energy
Density}) with respect to $\xi_\uparrow$ and  $\xi_\downarrow$
yields the system of two nonlinear equations for  $\xi_\uparrow$ and
$\xi_\downarrow$:
\begin{eqnarray} \label{eq:Steady State}
\tanh \frac{\xi_{\uparrow}-\xi_{\downarrow}}{2}\tanh \frac{\xi_{\uparrow}+\xi_{\downarrow}}{2}=h_0, \quad \nonumber \\
\frac{\sinh 2\xi_\uparrow+ \sinh
2\xi_\downarrow-2(\xi_\uparrow+\xi_\downarrow)}{(\cosh
\xi_{\uparrow}+\cosh\xi_{\downarrow})^2}=w.
\end{eqnarray}
The magnetic induction $\mathbf{B}=\langle \mathbf{h} \rangle$ is determined  by reduced magnetic induction $b=B/ 4\pi M_0$:
\begin{equation} \label{eq:Magnetic Flux}
b=-{1\over 2}{\partial  \mathcal {E}\over \partial
h_{0}}=\frac{(1+h_{0})\tanh\xi_\uparrow-(1-h_{0})\tanh\xi_\downarrow}{\xi_\uparrow+\xi_\downarrow}.
\end{equation}

Fig.~\ref{PhD}(a) shows the phase diagram  in the plane $w^2 -h_0$.
The area of the cryptoferromagnetic state is restricted by two lines
where the phase transition between the cryptoferromagnetic and the
Meissner states occurs:

1.  The line $\mathcal {E}=0$, which corresponds to the limit
$\xi_{\uparrow,\downarrow} \to \infty$. The values of $w$ and
$h_{0}$ on this line are connected by the relation $ w=1+h_0^2$.

2.  The line on which domains with magnetization opposite to the
external magnetic field vanish, $\xi_\downarrow=0$. The equation
describing this line is
        \begin{eqnarray}
  w_c=\sqrt{h_c}(1+h_c)  -{(1-h_c)^2\over 2}\ln{1+\sqrt{h_c}\over 1-\sqrt{h_c}},
        \end{eqnarray}
the critical size of the $\uparrow$-domain being
        \begin{eqnarray}
  \xi_{\uparrow c}=\ln{1+\sqrt{h_c}\over 1-\sqrt{h_c}}.
        \end{eqnarray}
The magnetic induction on the critical line is
        \begin{eqnarray}
b_c= 2\sqrt{h_c}\left(\ln{1+\sqrt{h_c}\over 1-\sqrt{h_c}}\right)^{-1}.
        \end{eqnarray}

Let us consider the left lower corner of this diagram where $w\ll $1
and $h_0 \ll 1$. The critical parameters on the line
$\xi_\downarrow=0$  as functions of $w$ are
\begin{equation} \label{eq:Critical Values: Particular Case}
h_{c}=\frac{1}{4}(3w)^{2/3}, \quad \xi_{\uparrow c} =(3w)^{1/3} .
\end{equation}
Aside from the critical line Eqs.~(\ref{eq:Steady State}) yield:
\begin{equation}
 h_0={\xi_\uparrow^2-\xi_\downarrow^2\over 4},~~
 w={\xi_\uparrow^3+\xi_\downarrow^3\over 3} .
 \label{eq:Particular Case}
\end{equation}
For small $h_0 \ll h_c$ one can solve
Eqs.~(\ref{eq:Particular Case}) analytically:
\begin{equation} \label{eq:Low Limit}
\xi_{\uparrow,\downarrow}=4^{1/3}\sqrt{h_c} \left(1- {h_0^2\over
4^{4/3} h_c^2}\right) \pm {h_0\over 4^{1/3} \sqrt{h_c}}.
           \end{equation}

\begin{figure}
  \includegraphics[width=0.4\textwidth]{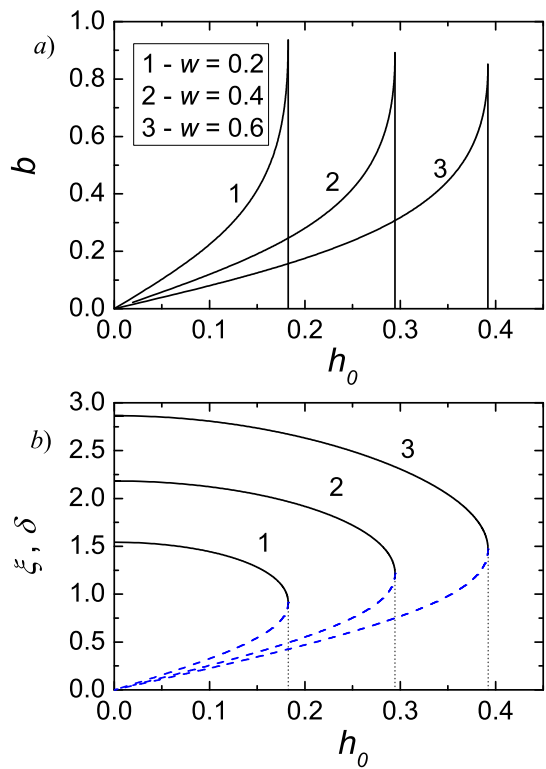}
\caption{\ (color online) ($a$) Magnetization curves and
($b$) magnetic field dependencies of $\xi=\xi(h_0)$ (solid line) and
$\delta=\delta(h_0)$ (dash line) are shown at different values of
parameter $w$. Vertical lines correspond to critical fields above
which the intrinsic domain structure collapses.}
\label{Magnetization Curves}
\end{figure}

Let  us consider the magnetization curve in the cryptoferromagnetic
state. The linear magnetic permeability is determined from two
relations connecting $\mu$ and $w$ with the period
$\xi=\xi_\uparrow+\xi_\downarrow \approx 2\xi_\uparrow$
\begin{equation}
\mu={db\over dh_0}={\coth \xi \over \xi},~~~w=\tanh \xi -{\xi \over \cosh \xi}.
           \end{equation}
In the limit $w\to 0$ the magnetic permeability is divergent: $\mu
\approx (2/3 w)^{2/3}$. The whole magnetization curves $b(h_0)$,
which were calculated numerically, are shown in
Fig.~\ref{Magnetization Curves}(a).

Fig.~\ref{Magnetization Curves}(b) shows the dependence of the
period $\xi=\xi_\uparrow+\xi_\downarrow$ and the difference of the
two domain widths $\delta=\xi_\uparrow-\xi_\downarrow$ as functions
of the reduced external magnetic field $h_0$.

Up to now we ignored the possibility of the transition to the mixed
state assuming that the first critical magnetic field $H_{c1}=\Phi_0
\ln \kappa  /4\pi \lambda^2$ essentially exceeds the characteristic
fields of the cryptoferromagnetic state. Here $\kappa=\lambda /\xi_0$ is the ratio of $\lambda$ to the coherence length $\xi_0$.  Let us now take into
account this possibility. Both the field $H_{c1}$ and the parameter
$w$ depend on the penetration depth $\lambda$, and it is
useful to introduce the reduced first critical field $h_{c1} =
H_{c1}/4\pi M_0 w^2=\Phi_0 M_0^3  \ln \kappa /\sigma ^2$, which does not depend on $\lambda$.
The reduced free energy of the mixed state with respect to the energy of the Meissner state is
\begin{eqnarray}
\mathcal{E}_m  = -\left(1+h_0  -{H^*\over 4\pi M}\right)^2
=-\left(1+h_0 -h_{c1} w^2\right)^2,~
                         \end{eqnarray}
where the field $H^*$ inside the mixed state differs from $H_{c1}$
by another logarithm factor, but we neglect it assuming  $H^*\approx H_{c1}$. The phase transition to
the mixed state may occur either from the Meissner state being
determined by the condition $\mathcal{E}_m=0$, or from the
cryptoferromagnetic state crossing the critical line on which  $\mathcal{E}_m=\mathcal{E}$. At zero
external field $h_0$ and small $w$  the phase transition between the mixed state and the
cryptoferromagnetic state occurs at $w_m \approx \sqrt{3}
/(2h_{c1})^{3/4}$. Thus whatever large  $h_{c1}$ is,  in the left
lower corner of the phase diagram  there is always {\em the
spontaneous vortex phase}, i.e., the mixed state without external
magnetic field.  The full phase diagrams at two finite values
$h_{c1}= $2 and 0.83 are shown in Figs. \ref{PhD}(b) and (c).
The cryptoferromagnetic state disappears from the phase diagram at $h_{c1}<0.5$.

Now let us analyze the phase transformations in the process of
cooling down below the SC critical temperature. This process is
different for spin and orbital SFs. In the case of spin SFs, when
the magnetic transition occurs at much higher temperature, one may
neglect temperature dependence of $M_0$, $\sigma$, and $h_{c1}$.
Then only $w^2 \propto 1/\lambda^2\propto \tau$ depends on relative
temperature difference $\tau=(T_c-T)/T_c$. On the phase
diagrams of Figs. \ref{PhD}(b) and (c) the state moves along
straight lines parallel to the horizontal axis $w^2$. From these
figures it is evident that just below the critical temperature the
system enters the mixed state. At further cooling down the system
crosses to the Meissner state either directly or through the area of
the cryptoferromagnetic state. For orbital SFs the cooling process occurs differently. In this case the
``magnetization'' $M_0 \sim \Phi /\lambda^2 \propto \tau$, and the
DW surface energy is a product of the condensation energy
$H_c^2(\tau) \sim [\Phi_0 /\lambda(\tau) \xi_0(\tau)]^2$ and the
coherence length $\xi_0(\tau) \sim \xi_0/ \sqrt{\tau}$: $\sigma \sim
\tau^{3/2} \Phi_0^2/\lambda_0^2 \xi_0$. Here $\lambda_0$ and $\xi_0$
are the penetration depth and the coherence length at zero
temperature. Then the parameters $w^2 \sim h_{c1}^{-1} \sim
(\lambda_0/\xi_0)^2$ do not depend on temperature whereas the
reduced magnetic field does: $h_0 =H_0/4\pi M_0 \propto 1/\tau$.
Thus in the field-cooling process the state moves along vertical lines
on the phase diagrams in Figs. \ref{PhD}(b) and (c). However,  as pointed out above, the cryptoferromagnetic state can
compete with the mixed state only if $h_{c1}$ is high enough. Since
$h_{c1} \sim (\xi_0/\lambda_0)^2$, this requires $\lambda_0$ not
large compared to $\xi_0$. According to \cite{Kirtley} the ratio of
$\lambda_0=190$ nm to $\xi_0=66$ nm is not too high indeed. But this means that the DW thickness  is not so small compared to $\lambda$ as assumed in our analysis.
Therefore for orbital SFs our analysis can provide only a qualitative but still credible picture of the phase
transformations.

In conclusion, we analyzed the conditions for appearance of intrinsic domains in  superconductors with broken time reversal symmetry, in which superconductivity coexists either with spin ferromagnetism or with ferromagnetism originated from the intrinsic angular momentum of Cooper pairs (chiral $p$-wave superconductors like $\mathrm{Sr_2 Ru O_4}$). Since these domains strictly speaking correspond not to a ferromagnetic but a globally antiferromagnetic state, the state was called {\em cryptoferromagnetic}. We considered competition of this state with the Meissner and the mixed states and found the phase diagram. This phase diagram can be checked with detailed measurements of magnetization curves.

This work has been supported by the grant of the Israel Academy of
Sciences and Humanities.

  \end{document}